\documentclass{elsart}

\usepackage{graphicx}

\usepackage{amssymb}

\def\R{\partial}
\newcommand{\bm}[1]{\mbox{\boldmath $#1$}}

\begin{document}

\begin{frontmatter}



\title{Curvature Effects on Surface Electron States in Ballistic Nanostructures}


\author{Hisao Taira and Hiroyuki Shima}

\address{Department of Applied Physics, Graduate School of Engineering, Hokkaido University,
Sapporo 060-8628 Japan
}

\ead{taira@eng.hokudai.ac.jp, shima@eng.hokudai.ac.jp}

\begin{abstract}
The curvature effect on 
the electronic states of a deformed cylindrical conducting surface
of variable diameter is theoretically investigated. 
The quantum confinement of electrons normal to the curved surface results 
in an effective potential energy that 
affects the electronic structures of the system at low energies.
This suggests the possibility that ballistic transport of electrons in 
low-dimensional nanostructures can be controlled by inducing a 
local geometric deformation.
\end{abstract}

\end{frontmatter}

\section{Introduction}
Recent advances in the manipulation of nanostructures have enabled the 
fabrication of reduced-dimensional quantum systems with novel geometry 
\cite{Tian,Lorke,Heinzel,Latg,Latil,Goker,Gridin,Qu,Chou,Sano,Pershin}. 
The understanding of their basic properties and the accurate modeling 
of their electronic structures are of vital importance for the 
manufacture of nanodevices and their applications. From the 
theoretical viewpoints, 
the peculiar features exhibited by quantum systems confined to 
low-dimensional curved geometry are of interest. These 
peculiarities arise
from an interplay between geometry and quantum physics; in fact, 
when the electron is strongly confined to a smoothly curved surface,
it experiences an effective potential energy
whose magnitude depends on  the local curvatures along the surface 
\cite{H. Jensen,R. C. T. da Costa,R. C. T. da Costa2,J. Goldstone,
M. Burgess,I. J. Clark,M. Encinosa,Entin,Entin2,M. Encinosa2,P. C. 
Schuster,Entin3,M. Encinosa3}. 
Due to this curvature effect, the electrons can not move around freely 
on the surface 
even in the absence of impurities or other interacting entities.
This implies that the quantum transport of 
low-dimensional nanostructures can be controlled by altering the local 
geometric curvature.

In the present work, 
we theoretically investigate the curvature effect on the
quantum properties of electrons
confined to a cylindrical surface. 
The local deformation of the cylindrical surface 
significantly affected the spatial profile of 
the electron eigenstates in the lowest-energy region.
This indicates the occurrence of a curvature-induced alteration in 
the ballistic electron transport in nanoscale cylindrical surfaces.


\section{The confining potential approach}

In quantum mechanics, the motion of quantum particles constrained 
to a two-dimensional curved surface is described by one of the two 
formalisms given below. One is the intrinsic quantization 
approach, in which 
the motion is constrained to the surface a priori; namely, a classical 
Hamiltonian is firstly constructed from coordinates and momentum intrinsic 
to the surface, following which the system is quantized canonically. The 
other is the confining potential approach, in which the particle 
is assumed to be confined by a strong force that acts normal to the 
curved surface. In this approach, the quantization of the motion 
perpendicular to the curved surface results in an effective potential 
that depends on the local surface curvature. Among the two formalisms, we 
employ the latter one since it offers a physically more realistic 
model of quantum confinement to curved surfaces. (In fact, in 
any real physical system, constrained motion is the result of a strong 
confining force.) It is mentioned that the confining potential approach 
was initially suggested by da Costa \cite{R. C. T. da Costa} 
; it has been successfully applied 
to quantum mechanical problems involving
novel geometries \cite{Cantele,Aoki,Koshino,Marchi,Fujita,Dandoloff}.

Let $(x^1, x^2, x^3)$ be a three-dimensional curvilinear coordinate.
This allows the parameterization of a curved surface of interest by
$\bm{r}=\bm{r}(x^1, x^2)$.
Then, a point $\bm{p}$ in this space can be determined through the relation
\begin{equation}
\bm{p} = \bm{r}(x^1, x^2) +x^3 \bm{n}(x^1, x^2),
\label{eq01}
\end{equation}
where $\bm{n}(x^1, x^2)$ is the vector normal to the curved surface. 
We now introduce a confining potential $V(x^1,x^2,x^3)$.
After a proper limiting procedure \cite{R. C. T. da Costa}, 
the potential becomes
$V=0$ if $x^3=0$ and $V=\infty$ otherwise. 
This allows us to separate the $x^3$ 
dependence in the Hamiltonian of the confined system and eventually 
provides the Schr\"odinger equation for curved surfaces:
\begin{equation}
-\frac{\hbar^2}{2m^*} \left[
\frac{1}{\sqrt{g}} \frac{\R}{\R x^i}
\left(\sqrt{g}g^{ij}\frac{\R}{\R x^j}\right)-
\left(H^2-D\right) \right] \sigma=E\sigma,
\label{eq02}
\end{equation}
where,
\begin{equation}
g_{ij} = \frac{\R \bm{r}}{\R x^i} \cdot \frac{\R \bm{r}}{\R x^j}, \quad
g={\rm det} [g_{ij}] \quad \mbox{and} \quad
g^{ij} = [g_{ij}]^{-1}.
\end{equation}
In Eq. (2), $\sigma(x^1, x^2)$ is the wave function of the confined particles, 
and $H(x^1, x^2)$ and $D(x^1, x^2)$ are the mean and 
Gaussian curvatures of the surface, respectively \cite{Riley}.
The occurrence of the non-trivial potential term, $-(H^2-D)$ 
in Eq.~(\ref{eq02}), is a direct consequence of the quantization 
of the motion normal to the surface. It is emphasized that
this potential term depends only on the local surface geometry
and not on the mass or charge of the particle.

 \section{Model and methods}
 \subsection{Schr\"odinger equations for deformed cylindrical surfaces}
%

\begin{figure}[ttt]
\begin{center}
\includegraphics[width=20pc]{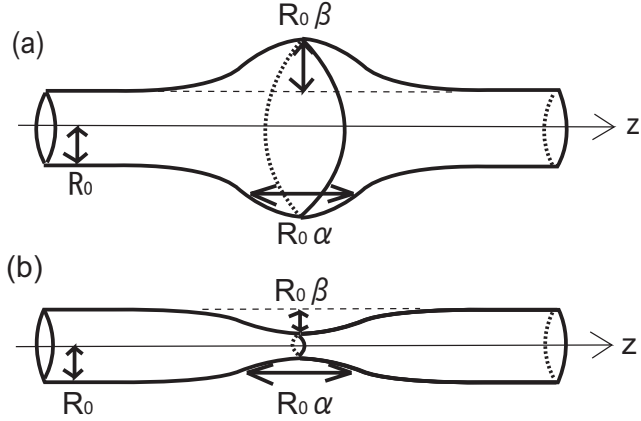}
\end{center}
\caption{Schematic illustration of a deformed cylindrical surface. 
The parameter $\alpha$ determines the spatial extent of the deformed 
part along the $z$-axis, 
and $\beta$ determines the extent of a bulge ($\beta>0$) or 
constriction ($\beta<0$).}
\end{figure}

Figure 1 gives a schematic illustration of a cylindrical 
surface subject to local deformation.\footnote{Similar cylindrical 
surfaces have been 
considered in Refs.~\cite{Cantele}, \cite{Marchi} and \cite{Parascandolo}.}
This curved surface is parameterized by
\begin{eqnarray}
\bm{r} = \bm{r}(z,\phi) = \left[ R(z)\cos\phi, R(z)\sin\phi, z \right],
\end{eqnarray}
where the radius of the cylinder is assumed to vary with $z$ as
\begin{equation}
R(z)=R_0\left[1+\beta\exp\left({\frac{-2z^2}{\alpha^2 R_0^2}}\right)\right].
\end{equation}
Hence, the geometry of the surface in question is determined by the 
two parameters: the parameter $\alpha$ determines the spatial extent 
of the deformed part along the $z$-axis, 
and $\beta$ determines the extent of a bulge ($\beta>0$) or 
constriction ($\beta<0$).

Our current objective is to deduce the electron eigenstates of the 
curved surfaces introduced above. It is noteworthy that, due to the 
rotational symmetry of the surfaces, the Schr\"odinger 
equation (\ref{eq02}) 
can be further simplified by means of the variable-separation method.
This is seen by substituting
\begin{eqnarray}
\sigma (z,\phi)=\frac{\eta(z)}{\sqrt{R(z)f(z)}}\frac{e^{im\phi}}{\sqrt{2\pi}},
\end{eqnarray}
into Eq.~(\ref{eq02}), with the definition
\begin{eqnarray}
f(z)=\frac{1}{\sqrt{1+(dR/dz)^2}}.
\end{eqnarray}
Then, we obtain the reduced 
Schr\"odinger equation for $\eta(z)$ as 
\begin{eqnarray}
-\frac{\hbar^2}{2m^*}\left[f(z)^2 \frac{d^2 \eta}{dz^2}+v(z) \eta(z) \right]
=\epsilon \eta(z),
\label{eq06}
\end{eqnarray}
where
\begin{equation}
v(z) 
= f(z)^2\left[ \frac{d\Gamma}{dz}-\Gamma(z)^2 \right]
+ H(z)^2-D(z)-\frac{m^2}{R(z)^2},
\label{eq07}
\end{equation}
and
\begin{eqnarray}
\Gamma(z) &=& -\frac{1}{2R(z)f(z)} \frac{d(Rf)}{dz}, \\
H(z)&=&f(z)^3\frac{d^2R}{dz^2}-\frac{f(z)}{R(z)}, \\
D(z)&=&-\frac{f(z)^3}{R(z)}\frac{d^2R}{dz^2}.
\label{eq08}
\end{eqnarray}
Consequently, the problem is reduced to solving the differential equatioin 
(8) with respect to $\eta (z)$. By observing the $\alpha$- and $\beta$- 
dependences of $\eta (z)$, we can clarify the curvature effect on the 
electronic structures of deformed cylindrical surfaces.

 \subsection{Tight-binding approximation}

In the actual calculations in Eq.~(\ref{eq06}), we have employed 
the tight-binding approximation; 
i.e., the continuous variable $z$ is discretized into a set of 
discrete numbers $i$ ($1\le i \le N$) with an equiseparation $a$. 
The resulting difference equation is 
\begin{equation}
t_0(2f_i^2-v_i)\eta_i-t_0f_i^2(\eta_{i+1}+\eta_{i-1})=\epsilon \eta_i ,
\end{equation}
where the constant $t_0 \equiv \hbar^2/(2m^* a^2)$ denotes the hopping 
energy between neighboring sites in an undeformed region. 
Nonzero surface curvature at the deformed region manifest itself in 
spatial modulation of $f_i$ and $v_i$ given in Eq.~(13). 
In fact, $f_i \equiv 0$ and $v_i \equiv \mathrm{const.}$ if the cylinder 
in question is flat (i.e., $R \equiv \mathrm{const.}$).

The differential equation (13) applies to real deformed cylindrical 
nanosurfaces in which the surface density of constituents (atoms or 
molecules) is homogeneous over the whole surface. 
In other words, the separation $r_{ij}$ of neighboring constituents 
needs to assume nearly constant so that the hopping energy remains 
to be constant. This condition, however, may violate when an 
external deformation against the original flat cylindrical surface is 
magnified. In the latter case, 
deformation-induced change in $r_{ij}$ can be taken into account by 
rewriting Eq.~(13) as 
\begin{equation}
t_0(2f_i^2-v_i)\eta_i-t_{ii+1}f_i^2\eta_{i+1}-t_{ii-1}f_i^2
\eta_{i-1}=\epsilon \eta_i .
\end{equation}
Here, $t_{ij}$ represents the spatially dependent hopping energy, 
which decreases exponentially with the separation $r_{ij}$ as
\begin{eqnarray}
t_{ij}= t_0 e^{-(r_{ij}-a)},
\label{eq10}
\end{eqnarray}
where
\begin{eqnarray}
r_{ij}=\int_i^jg_{11}(z)dz, \quad g_{11}(z)=\frac{1}{f(z)^2}.
\label{eq10}
\end{eqnarray}
Obviously, if $r_{ij} \equiv a$ for all $i$ and $j$, 
$t_{ij} \equiv t_0$ so that Eq. (14) reduces to Eq. (13).
In both the conditions mentioned above, the radius $R_0$ of an 
undeformed part of the cylindrical surface is set to $R_0=50$ 
and the length $L$ of the cylinder is 
$L=1000$ in units of $a$.



\section{Results and discussions}
\subsection{Localization length}
\begin{figure}[ttt]
\includegraphics[width=6.8cm]{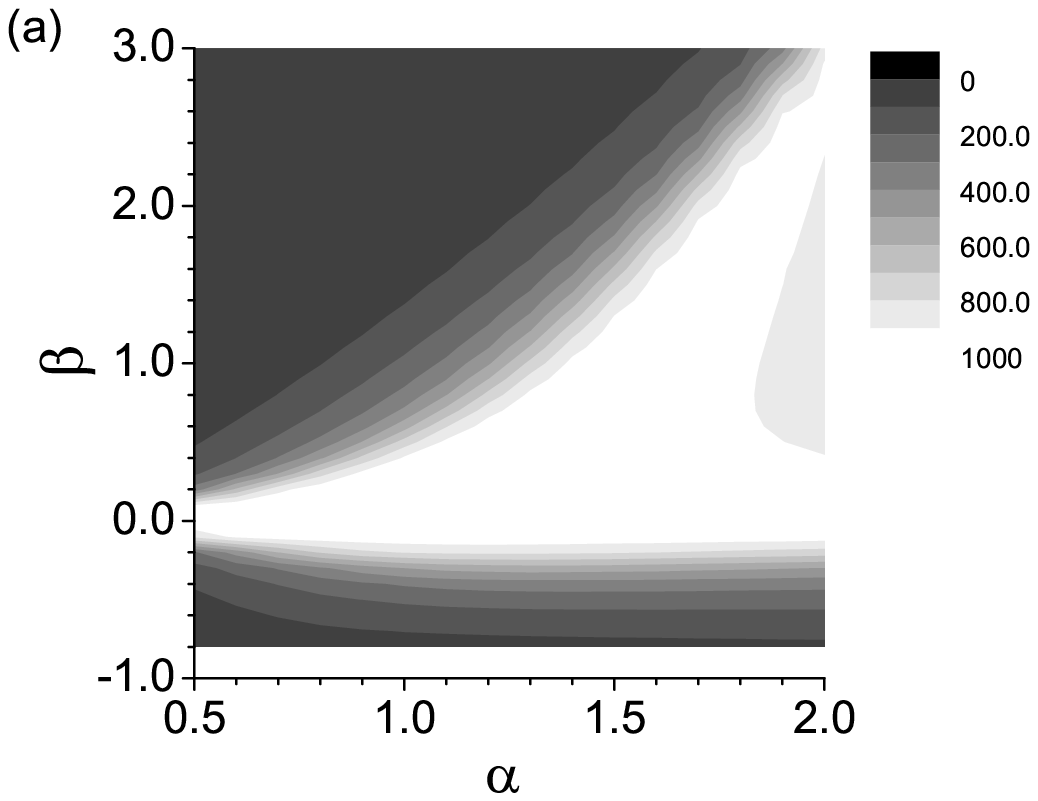}
\includegraphics[width=6.8cm]{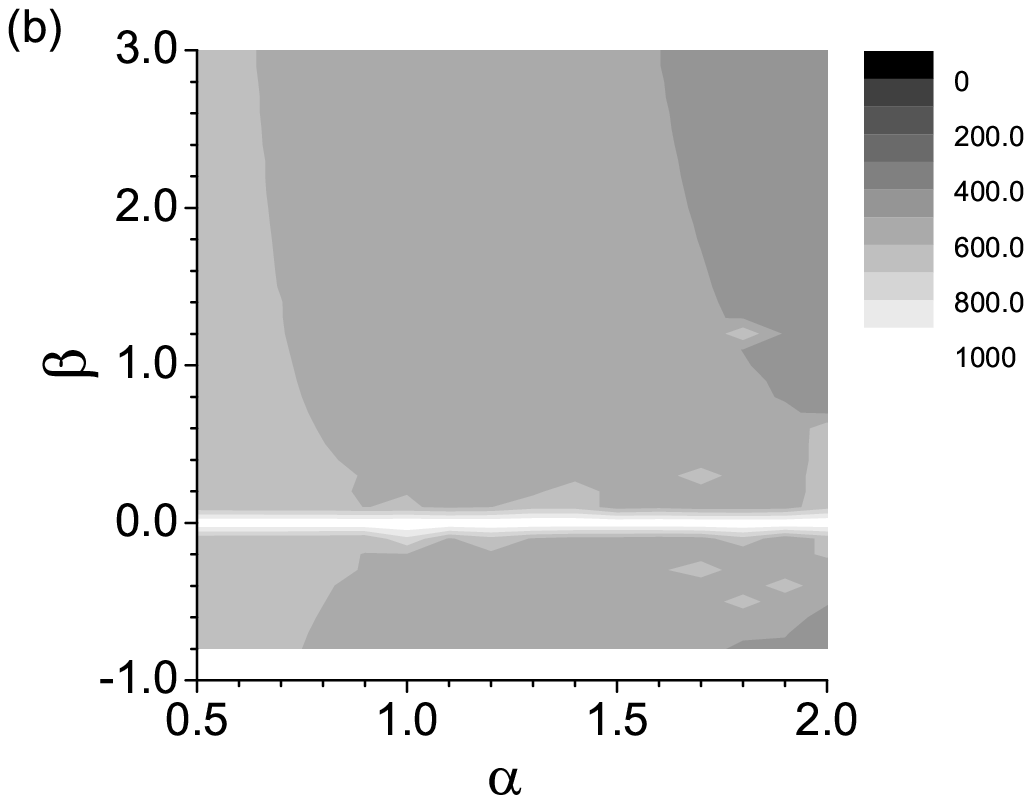}
\includegraphics[width=6.8cm]{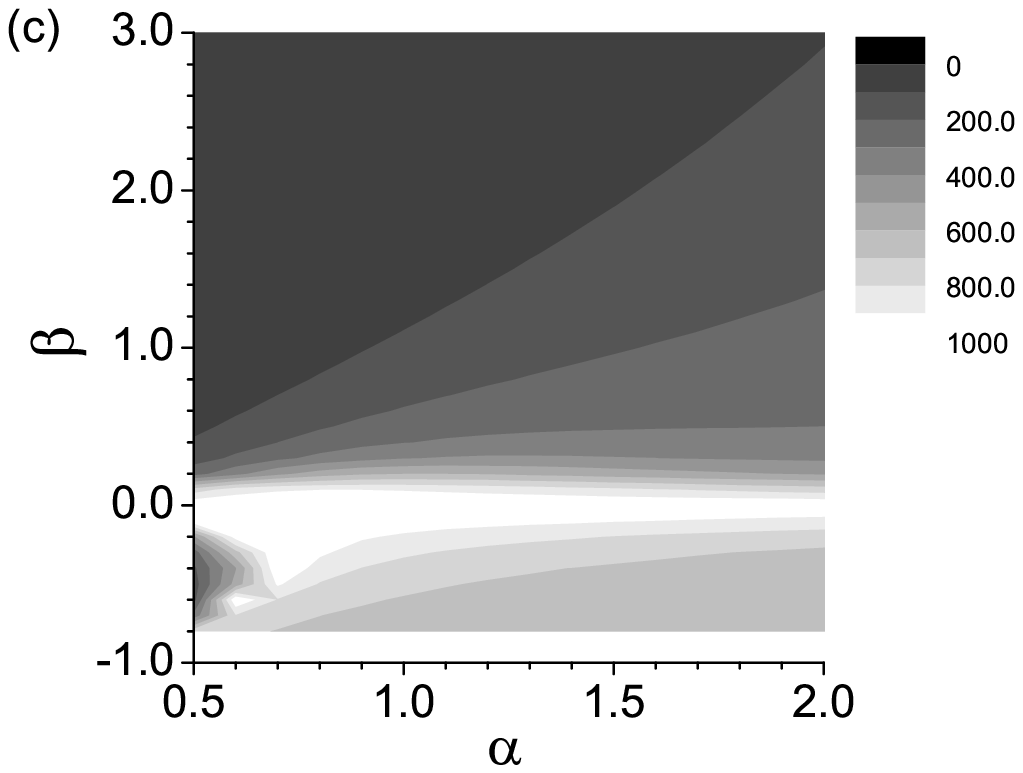}
\includegraphics[width=6.8cm]{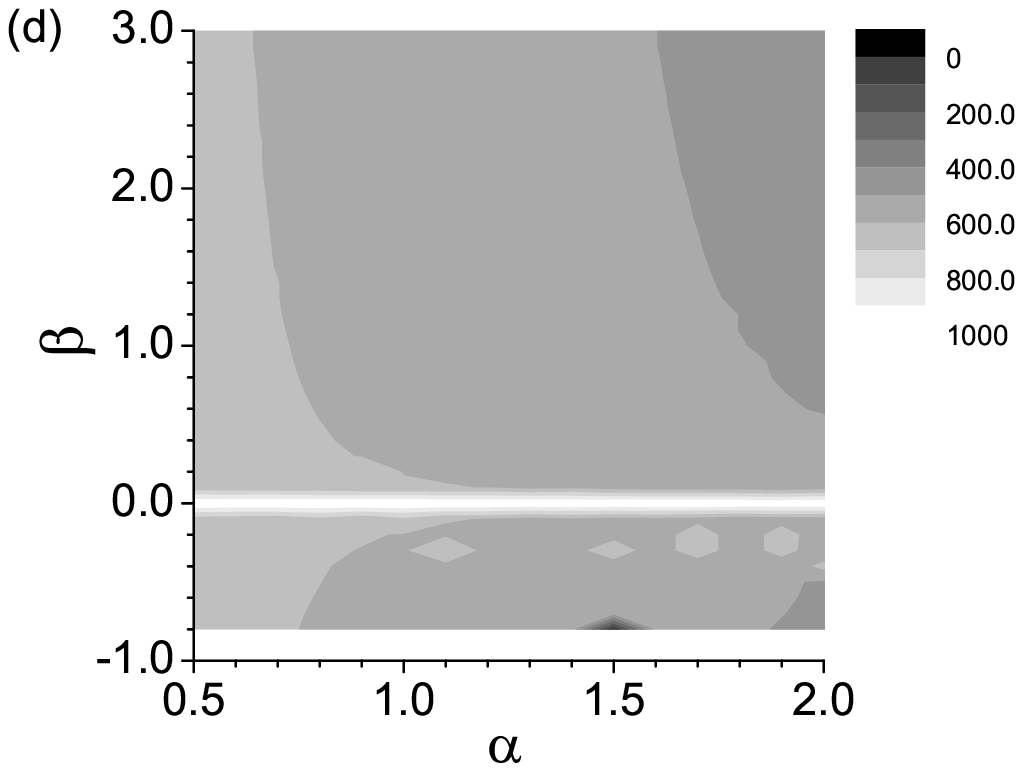}
\caption{Contour plots of the localization length $\xi$ 
of the lowest-energy eigenstate with (a)-(b) $m=0$ and (c)-(d) $m=1$.
In the two left panels, the hopping energy $t_0$ is assumed to be 
constant, while in the two right panels, the spatial 
variation of $t_{ij}$ is taken into account (see Eq.~(15)). 
The eigenstate is strongly localized in the dark region 
and extends over the entire system in the light region.}
\end{figure}
%

%
%
%
%
In order to clarify the curvature effect on the spatial profile 
of $|\eta(z)|^2$, we 
have calculated the localization length $\xi$ 
of the lowest-energy eigenstate defined by \cite{Fyodorov}
\begin{eqnarray}
\xi=\left| \sum_{i=1}^N \eta_i^4 \right|^{-1}.
\end{eqnarray}
This quantity provides a measure of the spatial extent of the wavefunction 
in question.\footnote{Instead of Eq.~(17), there is an alternative 
definition of localization length associated with the Lyapnov exponent; 
see Ref \cite{Shima} for example.} 
Hence, by examining the dependence of $\xi$ on 
the geometric parameters $\alpha$ and $\beta$, we obtain an understanding 
of the geometric effect on the electron system on  cylindrical surfaces. 

Figures 2(a)-(d) show the contour plots of the localization length $\xi$ 
of the lowest-energy eigenstate with (a)-(b) $m=0$ and (c)-(d) $m=1$.
In the two left panels, the hopping energy $t_0$ is assumed to be 
constant (see Eq. (13)), while in the two right panels, the spatial 
variation of $t_{ij}$ is taken into account (see Eq. (14)). 
Henceforth, we refer to the two models described by Eqs. (13) 
and (14) as models A and model B, respectively.

We see from the four panels that the ground-state eigenfunction 
may be strongly bounded
(dark region) or may extend over the entire system 
(light region) depending on the values of $\alpha$ and $\beta$. 
We first consider the result of model A; it is evident that 
in both the cases of $m=0$ (Fig.~2(a)) and $m=1$ (Fig.~2(c)), the 
$\alpha$ dependence of $\xi$
is completely different in the regions $\beta>0$ (bulging deformation)
and $\beta<0$ (constricted deformation).
With $m=0$, for instance, 
$\xi$ for $\beta>0$ tends to increase with $\alpha$ 
and eventually becomes equal to the system length $L$ at $\alpha \sim 1.5$. 
This indicates that the ground-state becomes extended ($\xi\sim L$) 
when reforming the bulging 
part to be more stretched in the direction of the $z$-axis.
On the other hand, $\xi$ for $\beta<0$ is almost invariant 
with regard to changes in $\alpha$,
and its value is always smaller than $L$.
As summarized, (i) the crossover behaviour of $\xi$ from the 
bounded to extended
states is observed only in the case of bulging deformation ($\beta>0$),
and within this range,
(ii) the magnitude of $\xi$ for a fixed $\beta$ rapidly increases at 
$\alpha\sim 1.5$.
These two findings 
suggest the possibility of controlling the ballistic electron transport in 
cylindrical nanostructures by introducing a subtle geometric deformation.

A similar crossover behaviour of $\xi$ was observed
in the case of $m=1$ (Fig.~2(c)). Nevertheless, in contrast to the 
case of $m=0$, the crossover occurs for negative $\beta$, not 
for positive. In fact, we have confirmed that 
when $m\ge 1$, the extended low-energy states occur only at $\beta < 0$, 
which is in contrast to the case of $m=0$, where the extended 
states occur only at $\beta >0$.
This difference as well as the global behavior of 
$\xi(\alpha, \beta)$ depicted 
in Figs.~2(a) and 2(c) can be accounted for by considering the 
contribution from each term
in the expression for $v(z)$ (see Eq.~(9)); 
detailed analyses will be presented elsewhere \cite{Taira,Taira2}.

While the localization length $\xi$ in the model A is responsive to 
the variations of $\alpha$ and $\beta$, that in model B exhibits 
only a slight dependence on the values of $\alpha$ and $\beta$. 
This is illustrated in Fig.~2(b) for $m=0$ and Fig.~2(d) for $m=1$. 
In both cases, $\xi$ decreases rather slowly with an increase 
$\alpha$ and $\beta$, which implies that a geometric deformation has 
almost no contribution to determining the spatial profiles of the 
eigenstates. 

\subsection{Spatial profile of the wavefunction}

\begin{figure}[ttt]
%
\includegraphics[width=6.8cm]{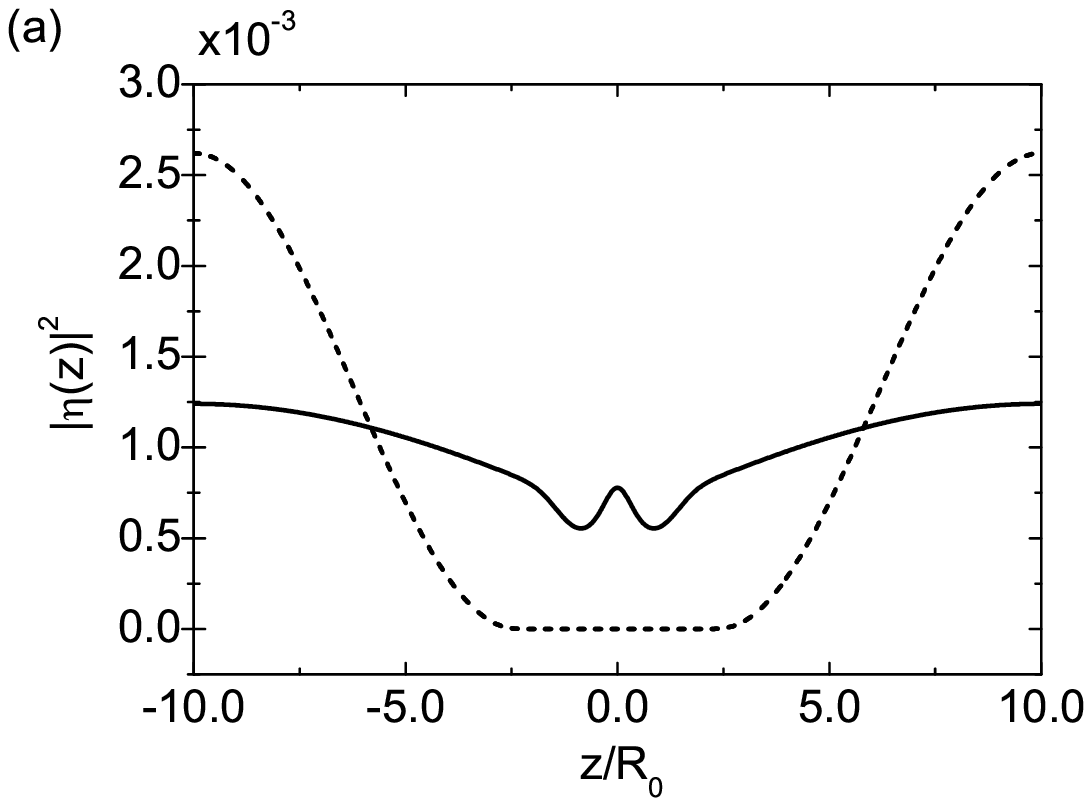}
\includegraphics[width=6.8cm]{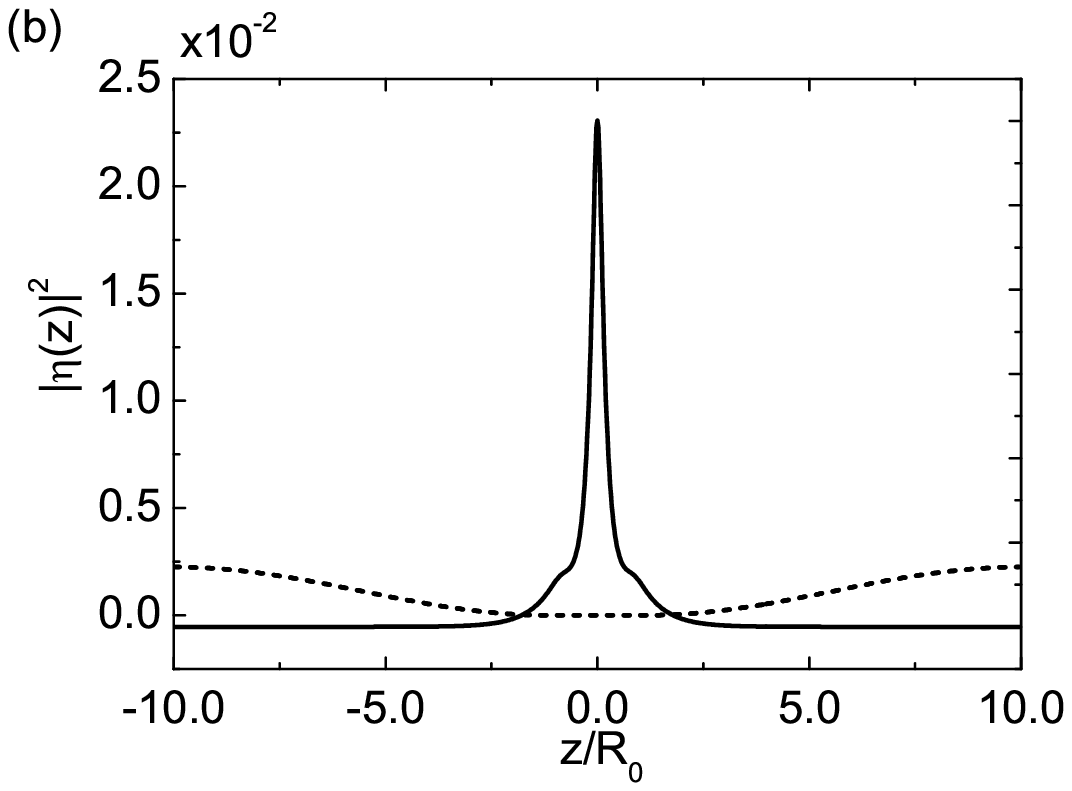}
\includegraphics[width=6.8cm]{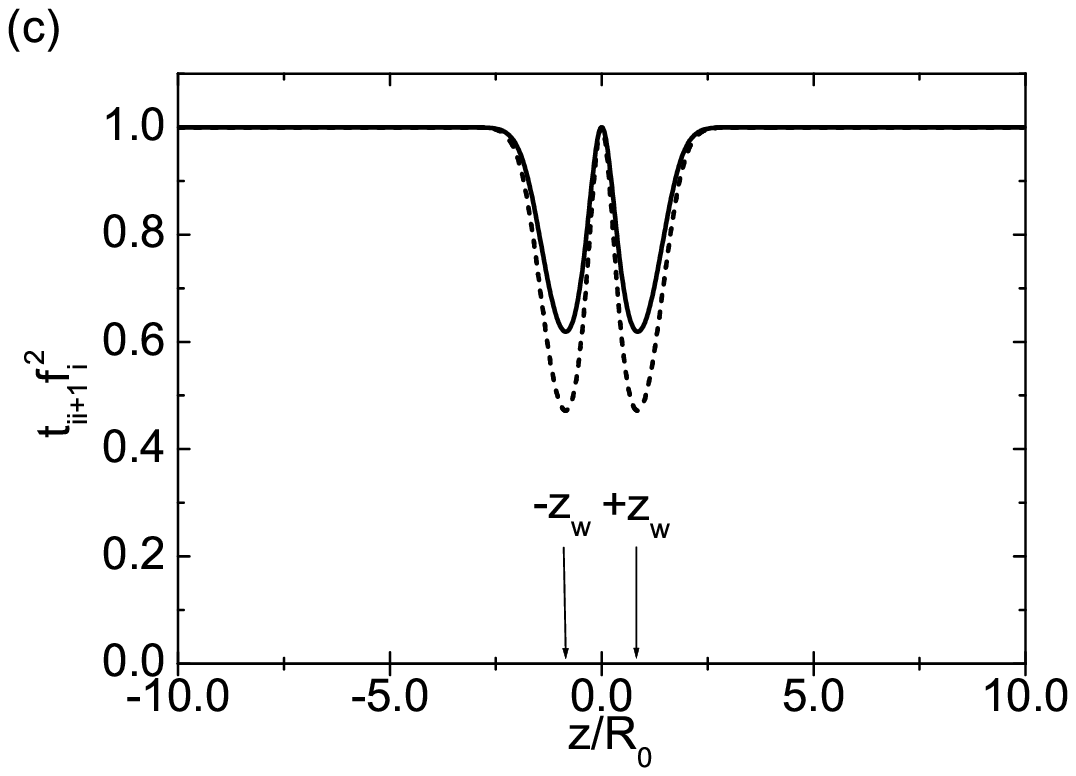}
\includegraphics[width=6.8cm]{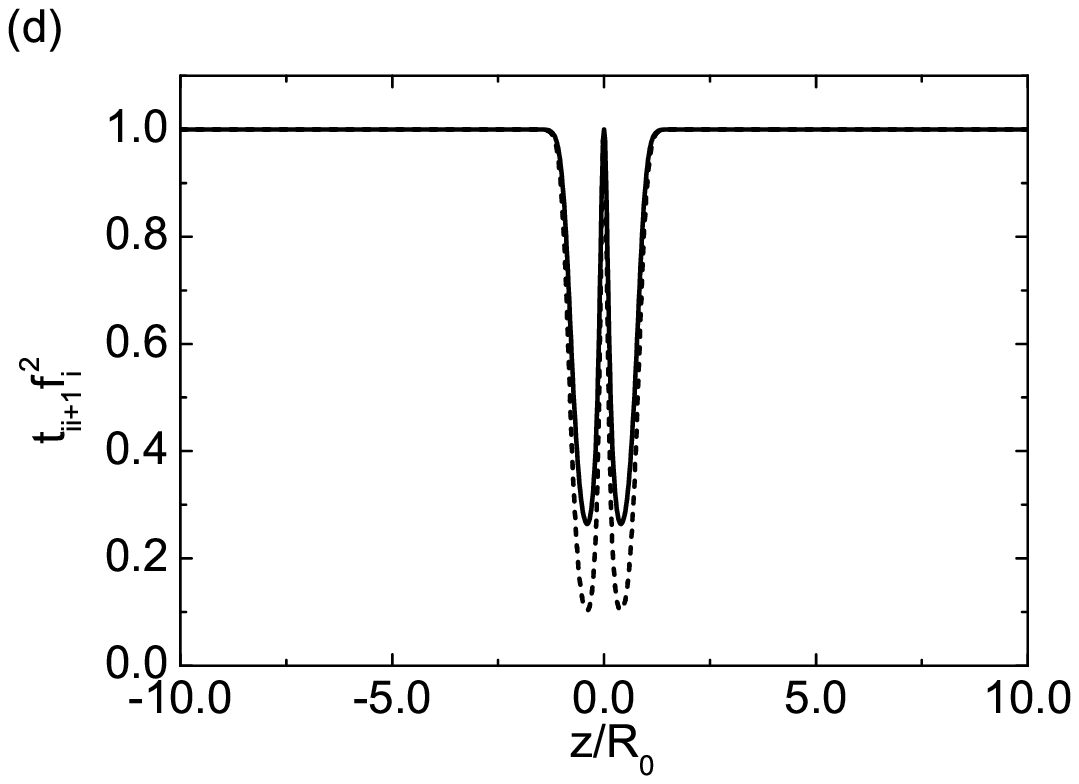}
\includegraphics[width=6.8cm]{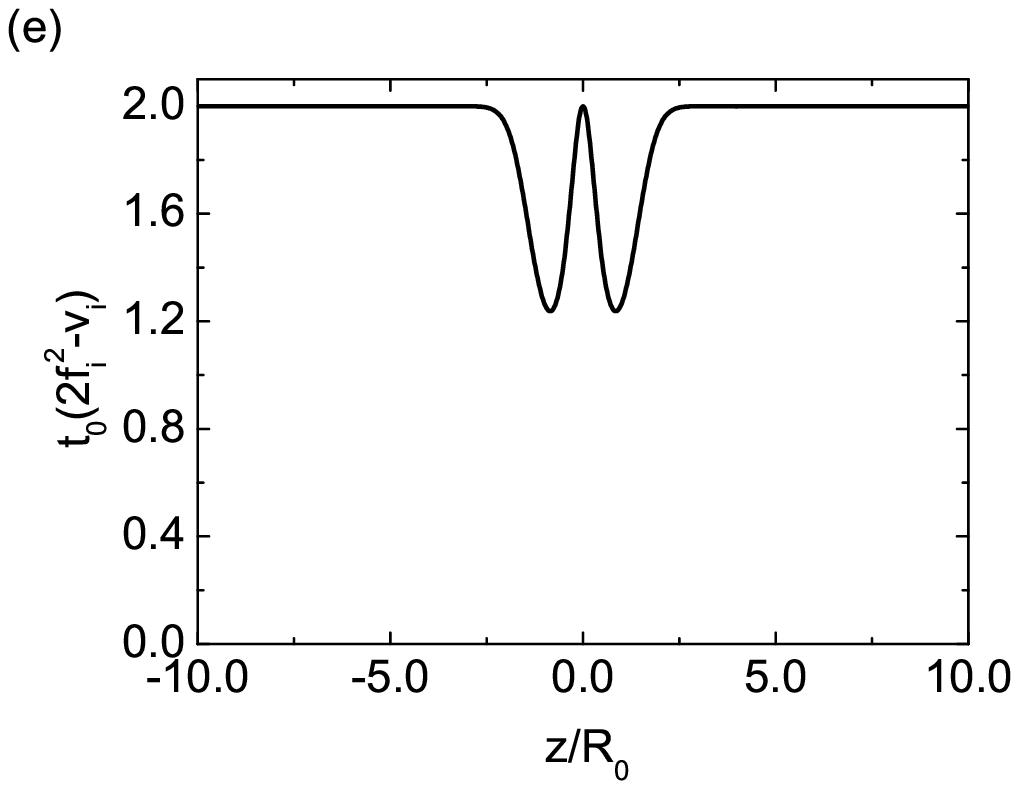}
\includegraphics[width=6.8cm]{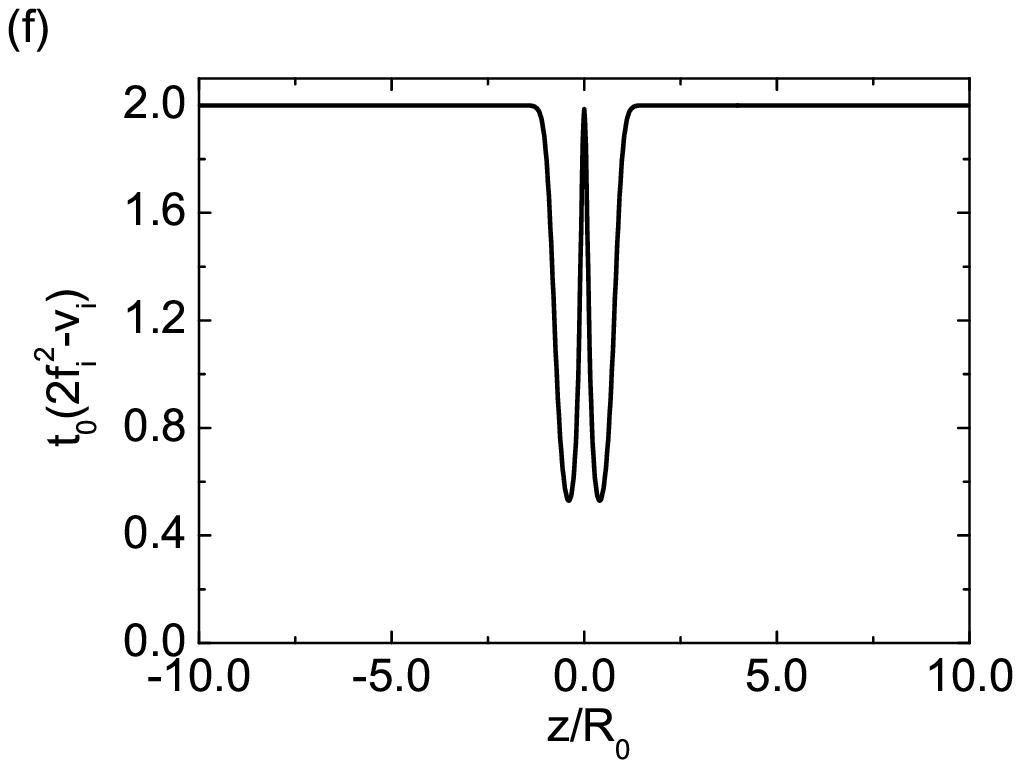}
\caption{The upper panels: Spatial profiles of $|\eta(z)|^2$ of the 
ground-state with $m=0$ for model A (the solid line) and for 
model B (the dotted line). The middle (lower) panels: The 
$z$-dependence of the off-diagonal (diagonal) element of the 
Hamiltonian matrix for model A (solid) and B (dotted). 
The parameters are set to be 
$(\alpha,\beta)=(0.8,1.0)$ in the left panels, and 
$(\alpha,\beta)=(1.7,1.0)$ in the right panels.}
\end{figure}
To understand the difference in the relevance of geometric 
deformation to the behavior of $\xi$ 
between in models A and B, 
we extract the spatial profile of the square amplitude of the ground-state 
eigenfunction for $m=0$ by fixing $\alpha$ and $\beta$ to be 
certain values. Figures 3(a) and 3(b) are plots of the 
profile of $|\eta(z)|^2$ for model A (the solid line) and for 
model B (the dotted line) with different values of 
($\alpha$, $\beta$). The parameters are set as 
$(\alpha,\beta)=(0.8,1.0)$ in the left panels and 
$(\alpha,\beta)=(1.7,1.0)$ in the right panels. We have also plotted 
the $z$-dependence of the diagonal and off-diagonal 
elements of the Hamiltonian matrix for the two models in Figs.~3(c)-(f), 
where $t_0$ is taken as units of energy.

We see from Figs.~3(a) and 3(b) that the wavefunction for model 
A is extended at $\alpha=0.8$, whereas it is spatially bounded around 
the deformed region ($z \sim 0$) at $\alpha=1.7$. Indeed, this behavior is 
completely consistent with  the contour plot of $\xi$ depicted in Fig.~2(a). 
In contrast, the wavefunction corresponding to model B exhibits almost no 
variance in its spatial profile for a different value of $\alpha$; 
it is fairly extended along the $z$-axis with a low (almost negligible) 
amplitude around the deformed region.

The disappearance of the bound state in model B is qualitatively 
understood by observing the $z$-dependence of the 
off-diagonal elements $t_{ij}f_i^2$, as shown in Figs.~3(c) and 3(d). 
In both plots, double-well structures whose bottoms are
located at $z=\pm z_w$ appear. It is noteworthy that the magnitude of the 
double well of 
model B is larger than that of model A. This is because 
in the model B, the value of the hopping 
energy $t_{ij}$ at $z=\pm z_w$ becomes smaller than $t_0$, 
since the separation $r_{ij}$ becomes larger than $a$ (see Eq.~(15)). 
Then, the quantum hopping across point $\pm z_w$ is slightly 
weak; thus, the 
wavefunction in model B tends to have a finite amplitude only 
within the undeformed region ($|z|>z_w$) in order to make the energy 
lower. As a consequence, the 
geometric deformation in 
model B neither engenders spatially bounded eigenstates as done 
in model A nor contributes significantly to the spatial 
profile of the lowest-energy eigenstates. This scenario indicates 
that the low-energy bound states observed in model A are difficult 
to realize by inducing an external mechanical deformation to 
a flat cylindrical conducting surface. 
We emphasize that the abovementioned findings regarding model B 
serve as a complementary result to the existing 
studies of the geometric effect on the electron transport 
in cylindrical surfaces, where the geometric conditions to realize 
the bound states were considered \cite{Cantele,Marchi}.

\section{Conclusions}
We have investigated the effect of local geometric deformation 
on the electronic states that are strongly confined to a thin 
cylindrical surface.
The spatial profile of the square amplitude of 
the lowest-energy eigenstates were numerically evaluated 
by the tight-binding approach; this was followed by the 
extraction the localization length of the eigenstate 
as a function of the geometric parameters $\alpha$ and $\beta$. 
The results indicate that, under certain conditions, a subtle 
geometric deformation could induce a drastic change in the
ballistic electron transport along nanoscale cylindrical surfaces.
We hope that our findings prove to be a fundamental basis for the 
development of quantum devices based on low-dimensional nanostructures.

\ack
We thank T.~Nakayama, K.~Yakubo and S.~Nishino for useful discussions.
This work was supported in part by a Grant--in--Aid for Scientific
Research from the Japan Ministry of Education, Science, Sports and Culture.
One of the authors (HS) is thankful for the financial support from 
the Murata Science Foundation. HS also acknowledges for the grant 
from the Mazda Foundation.
Numerical calculations were performed in part 
on the Supercomputer Center, ISSP, University of Tokyo.

\end{document}